\documentclass[unnumsec,webpdf,contemporary,large]{oup-authoring-template}%

\usepackage[subtle]{savetrees}
\usepackage{caption}
\graphicspath{{Fig/}}

\theoremstyle{thmstyleone}%
\theoremstyle{thmstyletwo}%
\theoremstyle{thmstylethree}%

\begin{document}

\journaltitle{}
\DOI{}
\copyrightyear{}
\pubyear{}
\access{}
\appnotes{Paper}

\firstpage{1}

\title[skandiver]{skandiver: a divergence-based analysis tool for identifying intercellular mobile genetic elements}

\author[1,2$\ast$]{Xiaolei Brian Zhang}
\author[1,2$\ast$]{Grace Oualline}
\author[3]{Jim Shaw}
\author[2,3,4\#]{Yun William Yu}

\authormark{Zhang and Oualline et al.}

\address[1]{\orgdiv{Department of Biological Sciences, Carnegie Mellon University}}
\address[2]{\orgdiv{Department of Computational Biology, Carnegie Mellon University}}
\address[3]{\orgdiv{Department of Mathematics, University of Toronto}}
\address[4]{\orgdiv{Department of Computer and Mathematical Sciences, University of Toronto at Scarborough}}

\corresp[$\ast$]{Equal contribution.}
\corresp[\#]{Corresponding author.}

\abstract{Mobile genetic elements (MGEs) are as ubiquitous in nature as they are varied in type, ranging from viral insertions to transposons to incorporated plasmids. Horizontal transfer of MGEs across bacterial species may also pose a significant threat to global health due to their capability to harbour antibiotic resistance genes. However, despite cheap and rapid whole genome sequencing, the varied nature of MGEs makes it difficult to fully characterize them, and existing methods for detecting MGEs often don't agree on what should count.
In this manuscript, we first define and argue in favor of a divergence-based characterization of mobile-genetic elements. Using that paradigm, we present \textsc{skandiver}, a tool designed to efficiently detect MGEs from whole genome assemblies without the need for gene annotation or markers. skandiver determines mobile elements via genome fragmentation, average nucleotide identity (ANI), and divergence time. By building on the scalable skani software for ANI computation, skandiver can query hundreds of complete assemblies against $>$65,000 representative genomes in a few minutes and 19 GB memory, providing scalable and efficient method for elucidating mobile element profiles in incomplete, uncharacterized genomic sequences.
For isolated and integrated large plasmids ($>$10kbp), skandiver's recall was 48\% and 47\%, MobileElementFinder was 59\% and 17\%, and geNomad was 86\% and 32\%, respectively. For isolated large plasmids, skandiver's recall (48\%) is lower than state-of-the-art reference-based methods geNomad (86\%) and MobileElementFinder (59\%). However, skandiver achieves higher recall on integrated plasmids and, unlike other methods, without comparing against a curated database, making skandiver suitable for discovery of novel MGEs.
~\\
~\\
\textbf{Availability}: \url{https://github.com/YoukaiFromAccounting/skandiver}
}

\maketitle

\begin{figure*}[t]
    \centering
    \includegraphics[width=0.9\textwidth,trim={0 20px 0 25px},clip]{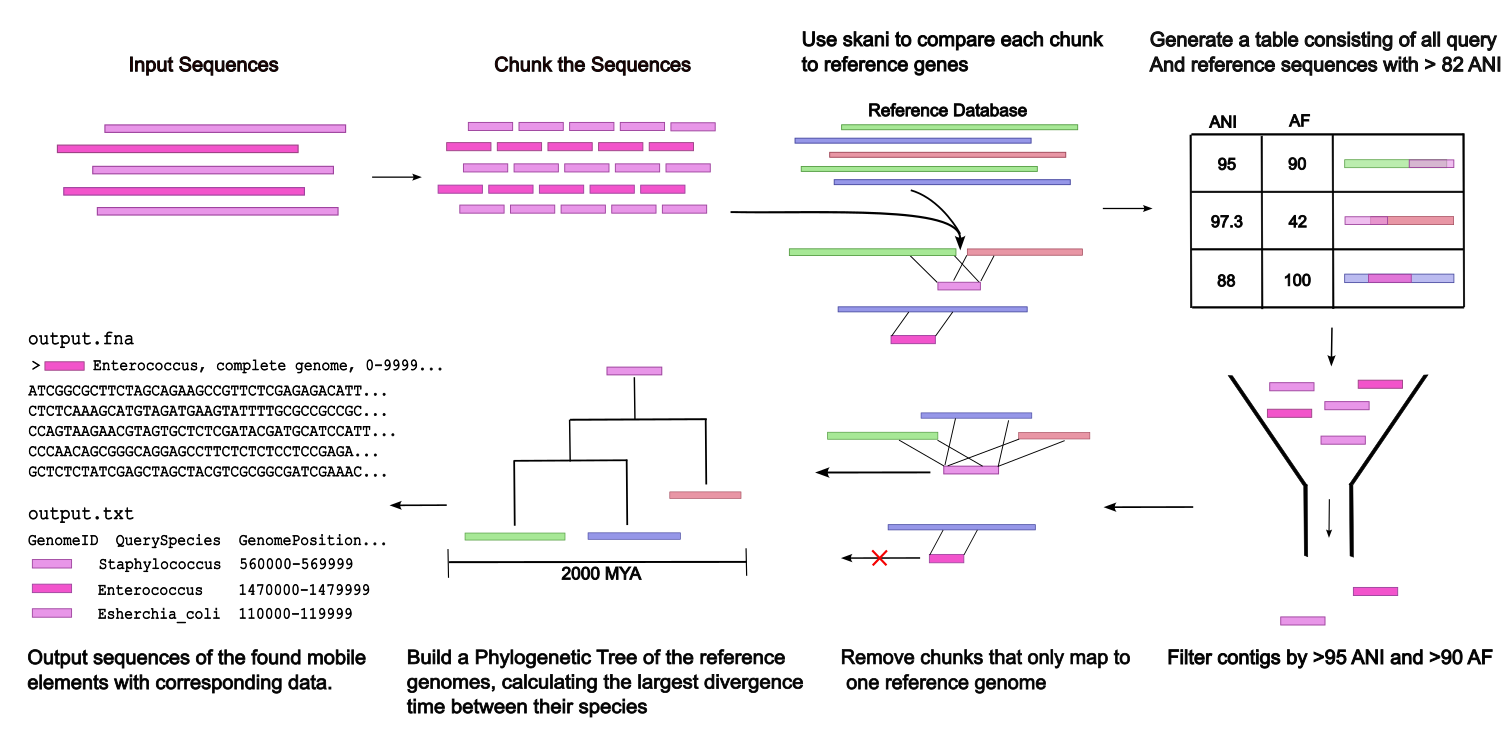}
    \caption{\textbf{A Composite Framework for Identifying Mobile Elements from Divergence Times.} skandiver analyzes whole genomes by pre-processing contigs into smaller genome fragments. Using skani, average nucleotide identity (ANI) and align fraction (AF) are found for each fragment. skandiver filters genome fragments by their ANI and AF to a set of representative genomes, and then performs a phylogenetic tree traversal to retrieve divergence times between every successful mapping to a representative genome. skandiver outputs the set of genome fragments that have mapped strongly to evolutionarily distant genomes, thereby linking these fragments with potential mobile genetic elements.}
    \label{fig:flowchart}
\end{figure*}

\begin{figure}[tbp]
    \centering
    \includegraphics[width=1\linewidth,trim={0 10px 0 0},clip]{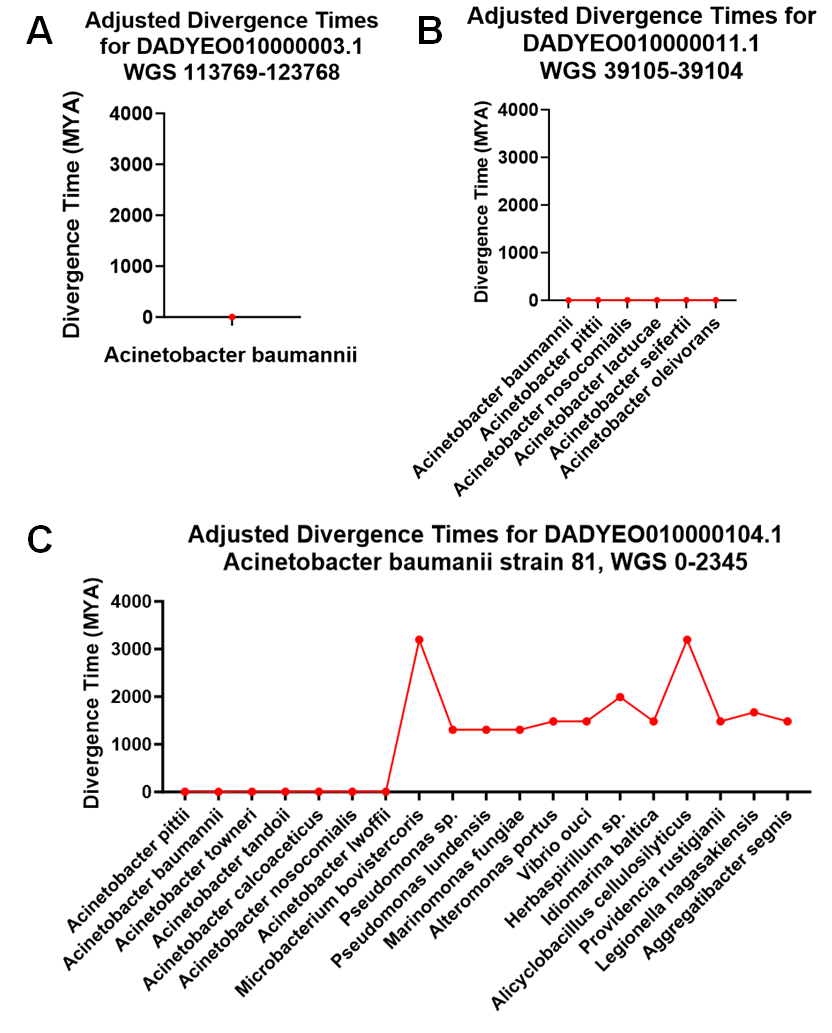}
    \caption{\textbf{Mobile Genetic Elements and Conserved Elements have distinct evolutionary divergence patterns.} Three separate genome fragments of \textit{A. baumannii} strain 81 (DADYEO010000104.1) were run in skandiver, and their phylogenetic mappings visualized: each point is a distinct match to a representative species of known divergence time from \textit{A. baumannii}. The nature of these genome fragments were verified using BLAST search and were characteristic of \textbf{A.} a random section of \textit{A. baumannii} genome not known to be mobile or conserved, \textbf{B.} a conserved region of \textit{A. baumannii} genome encoding ilvD dihydroxy-acid dehydratase protein, and \textbf{C.} an \textit{A. baumannii} plasmid X4-300 (CP064203.1).}
    \label{fig:divergence-idea}
\end{figure}

\section{Introduction}
Mobile genetic elements (MGEs) are snippets of sequences that can move around (either intra-cellularly within a genome, or inter-cellularly across species) independent of the host genomes \citep{shapiro2012mobile}. There exist many biological mechanisms associated with MGEs (viral phages, plasmids, transposons, etc.) but the end-results of especially inter-cellular trafficking of genetic material is often an accelerated evolutionary time-scale \citep{frost2005mobile}. 

To cite just one topical example, antibiotic resistance has been a critical challenge in healthcare, due to the overuse and misuse of antibiotics~\citep{durrant2020}. Antibiotic resistance genes thus contribute substantial fitness to the bacteria, and are strongly selected for by evolutionary pressures \citep{davies2010}. Bacteria often share genes through horizontal gene transfer~\citep{forster2022}, and it is thought that MGEs are instrumental in the ability of so-called ESKAPE pathogens to quickly develop resistance \citep{partridge2018mobile}. Identifying and detecting these MGEs is essential to gain insights into their formation and evolution, which may provide a foundation for developing preventative treatments to reduce the rise of antibiotic resistant strains ~\citep{johansson2021}.
Furthermore, the rapid evolution of these resistance genes highlights the need to detect MGEs without needing to reference a library of known contigs of antibiotic resistance.

Despite the importance of MGEs, there exist limitations with current bioinformatics software. One might quite reasonably hope that with the plethora of sequencing data now available to practitioners, we would be able to just throw data at the problem, but there are always trade-offs~\citep{berger2023navigating}. Many tools rely on gene-based classification methods, utilizing database searches and sequence homology in order to identify target sequences. One such tool, MobileElementFinder~\citep{johansson2021}, aligns assembled contiguous sequences (contigs) from a library of known mobile elements to input bacterial DNA sequences, in order to predict MGEs in those sequences. 
Another tool, geNomad, is a classification framework that can either use marker genes or a deep learning model to identify plasmids and viruses from gene annotations~\citep{camargo2023}.
However, both methods are limited to finding MGEs that are similar to previously discovered and annotated MGEs---MobileElementFinder explicitly through alignment, and geNomad implicitly through the machine learning model training set.
Furthermore, although fairly efficient, alignment and machine learning are both somewhat slower primitives compared to some of the more modern sketching-based sequence similarity techniques that we take advantage of in this manuscript~\citep{shaw2023}.

To address the limitation of prior tools being restricted to MGEs similar to already annotated ones, we target specifically intercellular MGEs from first principles using evolutionary divergence times.
Evolutionary divergence is typically measured in units of MYA (Million Years Ago), corresponding to our current best guess based on sequence analysis of proteins and genomes as to the timeframe of the last common ancestor between two species~\citep{doolittle1996determining}.
Our key insight is that mobile elements are by definition untethered to the background evolutionary process governing the rest of the genome---intercellular MGEs appear, mostly unchanged, in otherwise unrelated species.
We can thus predict that a sequence from a genome is a mobile element if the set of genomes it appears in have high mutual evolutionary divergence times, but that sequence does not appear in more closely related genomes.
Figure \ref{fig:divergence-idea} illustrates this idea graphically through an example from \textit{A. baumannii}. (a) Random sequences from a genome typically don't match any other species. (b) Conserved elements show up in many other species, but primarily only those are related (i.e. have low divergence times from the reference genome). (c) Mobile elements show up even in many distantly-related species with high evolutionary divergence times.

In this manuscript, we introduce our method \textsc{skandiver} for finding intercellular MGEs. Instead of using a previously annotated library of MGEs, skandiver measures the evolutionary divergence of the species a putative MGE has jumped between, enabling detection of MGEs database-based tools like geNomad and MobileElementFinder cannot.
skandiver is built around our prior tool skani~\citep{shaw2023} and the TimeTree of Life~\citep{kumar2022}. Skani is able to take a sequence and quickly and efficiently calculate average nucleotide identity (ANI) and the fraction of genomes aligned to one another (AF) of metagenomic sequences against a large dataset of genomes. This allows us to swiftly observe what chunks of sequences may map to multiple species of bacteria.
We utilize this by looking for sequences that have strong matches to a variety of bacterial species (as measured by TimeTree of Life evolutionary divergence times), indicating that it may be mobile in nature.

Using skandiver, we can detect MGEs in genomic sequences, and produce a summary table of their location, as well as the divergence time between the species that it mapped to. 
Researchers input assemblies, which are then fragmented and fed through skani, to find where those fragments may occur in the GTDB database~\citep{parks2022gtdb}. Then, fragments with strong matches to more than one species are taken, and the divergence time between those species is calculated to determine if that fragment is a conserved or mobile element. If the divergence time is low, it is likely a DNA sequence shared among species due to being essential. However, large divergence times point to the possibility of the fragment of DNA traveling between different species of bacteria. We output these MGE sequences and their corresponding data in an easy-to-read format, making it convenient for researchers with minimal coding experience. The speed and efficiency of utilizing this wrapper script with skani provides a versatile tool that operates with minimal memory for researchers to keep pace with the rapid trafficking of MGEs. Through its application, researchers will be able to quickly pinpoint sequences of DNA within bacterial genomes that are putative MGEs.

\begin{figure*}[bt]
    \centering
    \includegraphics[width=1\textwidth]{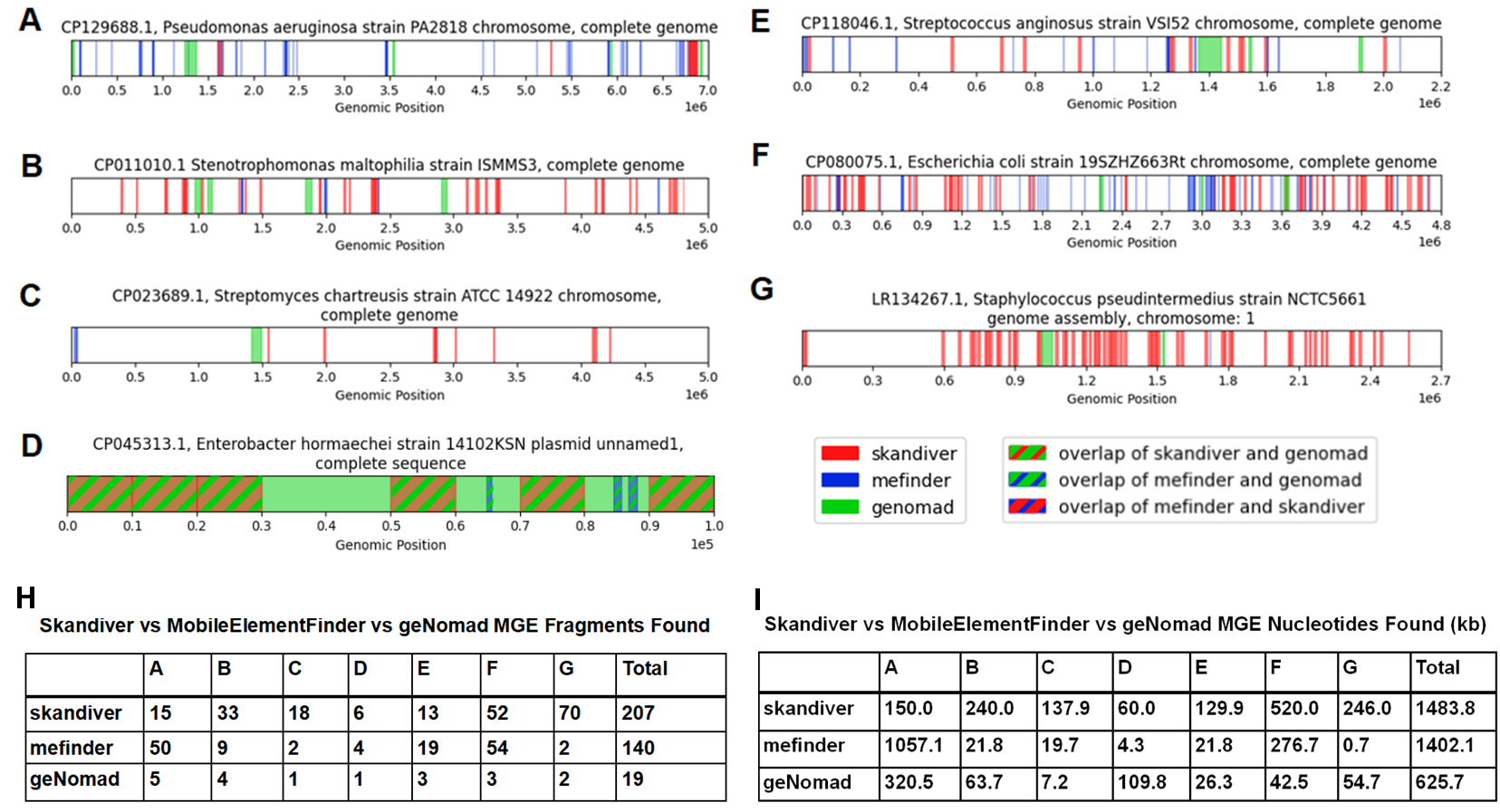}
    \caption{\textbf{Comparison of mobile element finding profiles of skandiver, MobileElementFinder (MEFinder), and geNomad: the three tools find different sets of putative genetic mobile elements.} Eight randomly selected contig-level whole genome assemblies were run through skandiver, MobileElementFinder, and geNomad. For each whole genome assembly, the regions corresponding to the potential mobile elements found by each method are labelled; \textbf{A.} \textit{Pseudomonas aeruginosa} strain PA2818 chromosome, \textbf{B. }\textit{Stenotrophomonas maltophilia} strain ISMMS3, \textbf{C. }\textit{Streptomyces chartreusis} strain ATCC-14922 chromosome, \textbf{D. }\textit{Enterobacter hormaechei} strain 14102KSN plasmid, stripes signify overlap between the methods, \textbf{E. }\textit{Streptococcus anginosus} strain VS152 chromosome, \textbf{F. }\textit{Escherichia coli} strain 19SZHZ663Rt chromosome, \textbf{G. }\textit{Staphylococcus pseudintermedius} strain NCTC5661. \textbf{H. } The amount of non-merged regions deemed a potential mobile element by each method. \textbf{I. } The amount of nucleotides deemed part of a potential mobile element by each method.} 
    \label{fig:comparison-barcodes}
\end{figure*}

\section{Materials and Methods}

\subsection{Algorithm Design}

The methodology behind using skandiver to identify potential mobile genetic elements is outlined in Figure~\ref{fig:flowchart}. The process can be divided into four main steps: 

\begin{enumerate}
    \item \noindent \textbf{Fragment genomes}: Genome assemblies of interest are fragmented/chunked into smaller regions, typically regions 10000 base pairs in length, depending on the granularity of specific genomic elements to be identified---in this manuscript, all our results are with 10kbp chunks for technical reasons related to skani's default implementation, so smaller MGEs may be missed.
    \item \noindent \textbf{Skani search against a database of representative genomes}: Fragmented genomes are searched against a database of representative genomes using skani, a robust and accurate ANI comparison tool.
    \item \noindent \textbf{Calculate divergence times}: skani search results are filtered to only high ANI and AF; we then annotate significant matches with evolutionary divergence times using the TimeTree of Life~\citep{kumar2022}.
    \item \noindent \textbf{Predict genomic elements}: we analyze patterns within skani search results and divergence times to output a set of potential mobile genetic elements, with genomic position and divergence time information. 
\end{enumerate}

\subsection{Augmenting skani search for genome fragments}

For whole genome assemblies and other similar metagenome-assembled genomes (MAGs) for which genomes may be contaminated or incomplete, skani search accurately and efficiently computes ANI between a query genome and a reference genome~\citep{shaw2023}. It does this by chaining a set of exact non-overlapping k-mer matches between the query and reference genomes to estimate the ANI. However, skani is designed for computing global similarity, rather than local similarity.
For MAGs with contigs that are large ($>$150,000bp), skani search will only return a hit if there is sufficiently high ANI and alignment between the entire reference and query. For example, when searching 5 whole genome contig-assembled query assemblies of \textit{Enterococcus faecium} against the Genome Taxonomy Database’s (GTDB)~\citep{parks2022gtdb} representative genomes as reference, skani search returns either zero matches between the query genome and the representative genome, or returns one match to the assembly’s original species (\textit{E. faecium}).

For finding MGEs, we are more interested in local similarity, so we augmented skani search to focus on genome fragments.
We thus introduced a pre-processing step in which MAGs are fragmented into non-overlapping chunks of sequence (10kbp in length by default). This step aimed to optimise the sensitivity of the skani search by providing a segmented representation of overall genomic content within metagenomic datasets, though it does restrict skandiver to only finding MGEs that are $>$10kbp---also, because of edge effects related to the overlap between chunks and the MGEs, $>$20kbp length MGEs are needed to guarantee that at least one chunk comprises only MGE sequence. Fragmentation was accomplished using a simple Python script that utilises the SeqIO and SeqRecord modules from the Biopython library~\citep{cock2009biopython}; the script takes as input a directory containing genomic sequences in fasta format, a desired output directory, and the preferred fragmentation size in nucleotides. The resulting fragments are then concatenated into a singular multifasta file, preserving original genomic identification information as well as fragment location within the genome.

Following genome fragmentation, repeated skani searches of 5 whole genome contig-assembled query assemblies of \textit{E. faecium} against the GTDB representative genome database returned over 1500 matches of over $90\%$ ANI between the query contig fragments and the reference database. More importantly, rather than returning only matches to a query’s original species, fragments now matched to a diverse array of bacterial species. For example, fragment 40581-42248 of \textit{$NZ\_JADVBE\textit{010000002.1}$} \textit{Enterococcus faecium} strain VRE32783 contig 00019, which previously was undetected by plain skani search, exhibited matches to \textit{Streptococcus pasteurianus}, \textit{Blautia argi}, \textit{Dilemma fastidiosa}, \textit{Eubacterium ramulus}, and many more reference bacteria following genome fragmentation. This is of particular interest for the detection of mobile genetic elements, many of which are insertion sequences or plasmids less than 10kbp in size ~\citep{khedkar2022}. 

\subsection{Filtering for Potential Mobile Genetic Elements}

ANI has previously been used to identify insertion sequences between 70 bp to 200 kbp by MGEfinder, which required query sequences to share at least 98.5$\%$ ANI with reference sequences to be classified as a mobile genetic element~\citep{durrant2020}. However, as our analysis method aims to directly identify mobile genetic elements from the query sequences themselves without the use of reference MGEs, we used a slightly more lenient requirement of at least 95$\%$ ANI and 90$\%$ align fraction (fraction of the query genome that aligns to the reference genome)---intuitively, two instantiations of an MGE that would have both mapped to the same reference sequence can be further apart from each other.

Another justification for our thresholds comes from 
prior work for identification of uncultivated virus segments---using the Minimum Information about any (x) Sequence (MIxS) standard, \cite{roux2019} suggested thresholds of 95$\%$ ANI and 85$\%$ alignment fraction for identification of virus segments. Our ANI threshold was thus in line, though we use a slightly higher align fraction threshold. All skani search results were extracted by using the awk Linux command and converted to comma-separated value file format following filtering for ANI and align fraction threshold.

\subsection{Estimating Divergence Time}

Divergence time between reference and query species was determined using TimeTree of Life, a public database of evolutionary and divergence times between major clades of organisms~\citep{kumar2022}. Using the GTDB Taxonomy Browser resource, we created a list of the 20,000 most common bacteria taxons. A Newick file representing a tree of divergence times for these bacteria taxons was then generated using the TimeTree database. Importantly, taxa for which divergence time was estimated with considerable uncertainty or for which discrepancies arise from individual time estimations were labelled according to their ``adjusted divergence time'' rather than their ``median divergence time''. This process was repeated until the complete timetree of GTDB common species was built.

Theoretically, potential mobile genetic elements should exhibit a distinct pattern of divergence times relative to their reference species, as we illustrated in Figure \ref{fig:divergence-idea}.
One might be tempted to try to characterize this pattern by considering evolutionary fitness and likelihood of mutations: some MGEs, such as those for antibiotic resistance genes, increase fitness relative to random segments of DNA, and certainly a large fraction of MGEs have coding regions within them.

However, there exists a much simpler characterization that we can perform based on recency of horizontal movement: a sequence that is shared by two species with high divergence times either must have remained largely unchanged over millions of years of evolution, or have horizontally moved more recently, which is the very definition of an intercellular MGE.
This phenomenon should be represented by matches between a query species and reference species with high divergence times.
One way to easily further distinguish between conserved elements and mobile elements is by whether the sequence is also present in other closely related species, since if it is not, then it likely isn't a conserved element.
Fortunately, this is rarely necessary, since we are operating at the genetic level---even highly conserved proteins show substantial changes at the genetic level after sufficiently many million years of evolution, so conserved element will have an average divergence time that is much lower than that for mobile elements.

This theoretical framework guided the validation of our method (Figure \ref{fig:divergence-idea}). We ran skandiver on five whole genome assemblies of \textit{Acinetobacter baumannii} to determine if MGEs can be distinguished by their pattern of divergence times. Divergence times were manually extracted from skandiver results and visualised using GraphPad Prism 10. 

We then manually extracted a few representative genome fragments, showcasing the divergence pattern of random sequence, a conserved element, and a mobile element. Characteristics of query sequences were verified using~BLAST \citep{altschul1990basic} search and plasmid/viral databases IMG/PR~\citep{camargo2024mode} and IMG/VR~\citep{paez2016img}. As shown in Figure \ref{fig:divergence-idea}, where \ref{fig:divergence-idea}A represents a random genome fragment, \ref{fig:divergence-idea}B represents a fragment corresponding to conserved protein ilvD dihydroxy-acid dehydratase (verified via BLAST search), and \ref{fig:divergence-idea}C represents a fragment corresponding to the mobile genetic element plasmid X4-300 (CP064203.1), three distinct patterns of divergence can be observed.
\begin{itemize}
    \item In Figure \ref{fig:divergence-idea}A, the random fragment simply matches to its original species with divergence time of 0, which is not indicative of a conserved or mobile element. 
    \item In Figure \ref{fig:divergence-idea}B, the conserved fragment matches to many species within the same Acinetobacter clade, with divergence times ranging from 0 to 5 MYA, indicating of a conserved element residing within a single bacterial clade due to shared ancestry.
    \item In Figure \ref{fig:divergence-idea}C, the MGE matches to both many species within the Acinetobacter clade and also a wide range of diverse species with divergence times ranging from 1000 to 3000 MYA. The broader range of divergence times supports the notion of a mobile genetic element capable of horizontal movement across distinct and diverse bacterial clades over an extended evolutionary timeline.
\end{itemize}
These distinctive evolutionary patterns within genomic elements are what skandiver uses to differentiate between conserved and mobile genetic elements within an assembly.

\subsection{Benchmarking Details}
We compared skandiver against MobileElementFinder~\citep{johansson2021} and geNomad~\citep{camargo2023}---although skandiver is not fully comparable to these other methods because skandiver does not rely on a curated database, these seemed like the closest possible comparisons. Default detection settings were used for both MobileElementFinder and geNomad. In skandiver, skani search/dist was run using c = 150, m = 1000, t = 10. The genome fragmentation length was set to 10000 bp, but can be adjusted depending on the expected mobile element size. 
Benchmarks were evaluated on a Dell Poweredge 660xs 1U server with two 16-core Xeon 5416S processors and 512 GB memory using 10 threads. 

\section{Results and Discussion}

\subsection{skandiver, MobileElementFinder, and geNomad find different putative mobile elements}
We analyzed 7 different contig-level whole genome assemblies and highlighted regions identified as putative mobile elements (Figure \ref{fig:comparison-barcodes}). 
skandiver and MobileElementFinder found comparable amounts of putative mobile elements---207 fragments of 1483.8kb total for skandiver and 140 fragments of 1402.1kb total for MobileElementFinder. geNomad found somewhat fewer---19 fragments of 625.7kb nucleotides total, though it should be added that geNomad is trained to predict a subset of the various types of mobile elements.
However, more interesting, with one major exception, the putative mobile elements found are very different, almost disjoint in some cases, with very little to no overlap between methods (Supplementary Figure S1). The only notable exception was in Figure \ref{fig:comparison-barcodes}D, where the entire contig we analyzed was itself a plasmid---geNomad marked the entire plasmid as a MGE, whereas skandiver and to a lesser extent MobileElementFinder marked overlapping chunks hits.
Thus, it seems that the three different approaches represented here---divergence for skandiver, reference database alignment for MobileElementFinder, and machine learning on a database for geNomad---are orthogonal to each other, despite having similar aims. 
From these discordant results, we conclude that none of the methods are fully capturing the wide breadth of mobile genetic elements in nature.

\begin{figure}[hbt!]
    \centering
    \includegraphics[width=1\linewidth]{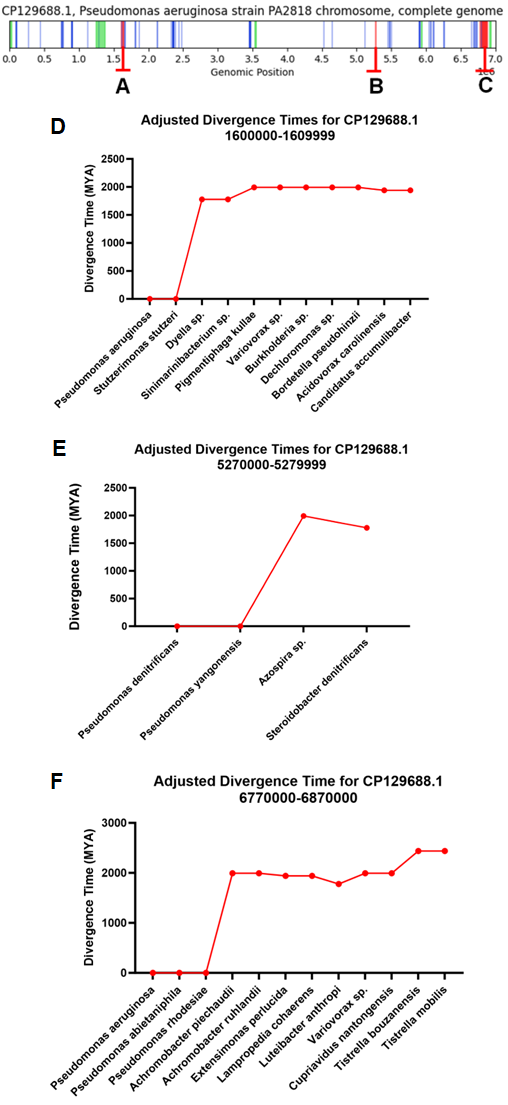}
    \caption{\textbf{skandiver finding \textit{Pseudomonas aeruginosa} plasmid p12CC3$\_$GES-5 previously elusive to geNomad and MobileElementFinder within \textit{Pseudomonas aeruginosa} strain PA2818 chromosome assembly.} \textbf{A. }, \textbf{B. }, and \textbf{C. } correspond to regions of CP129688.1 designated as potential mobile elements by skandiver. \textbf{D. }Visualization of species matches for fragment 1600000-1609999 found by skandiver (overlaps with fragment 1612765-1614092 found by MEFinder). \textbf{E. }Visualization of species matches for fragment 5270000-5279999 found by skandiver (does not overlap with MEFinder or geNomad). \textbf{F. }Visualization of species matches for fragment 6770000-6870000 found by skandiver (overlaps with fragment 6787235-6788562 found by MEFinder).}
    \label{fig:barcode-justification}
\end{figure}

\subsection{skandiver can find novel mobile genetic elements that other methods miss}
For further validation of skandiver's hits, we do a deep dive into Figure \ref{fig:comparison-barcodes}A in Figure \ref{fig:barcode-justification}. We chose this example because skandiver only had 3 substantial hits, so we can do an exhaustive analysis without any cherry picking.
Analyzed here were a set of MAGs from \textit{Pseudomonas aeruginosa}. While there was very little overlap between the three methods, skandiver and MobileElementFinder were both able to find a potential mobile genetic element at around 1.6 Mbp into a whole genome assembly of \textit{P. aeruginosa} (CP129688.1) (Figure \ref{fig:barcode-justification}A,D). Upon performing BLAST nucleotide search of this fragment against the NCBI Database~\citep{wheeler2007database}, it was revealed that this fragment shared 100$\%$ query coverage and 99.97$\%$ percent identity with the \textit{P. aeruginosa} plasmid 2017-45-85 (CP109756.1) Similarly, there was another potential mobile genetic element at around 6.7 Mbp that was also found by both MobileElementFinder and skandiver (Figure \ref{fig:barcode-justification}C,F). 

More interestingly, skandiver was able to find a novel MGE that was previously uncharacterized by both MobileElementFinder and geNomad at approximately 5.27 Mbp into the same whole genome assembly (Figure \ref{fig:barcode-justification}B,E). We were able to visualise this potential mobile element’s sharedness between species by plotting the adjusted divergence time in MYA against the different species the fragment mapped to following skani search. As seen in Figure~\ref{fig:barcode-justification}E, this fragment of interest as determined by skandiver mapped strongly to both \textit{Pseudomonas dentrificans} and \textit{Pseudomonas yangonensis} with negligible divergence times, but also to \textit{Azospira sp.} and \textit{Steroidobacter dentrificans} with significant divergence times. We verified that the fragment has elements of mobility by performing a BLAST Search against the NCBI Database, which revealed that this particular fragment shared 98$\%$ query coverage and 94.071$\%$ percent identity with the \textit{Raoultella planticola} plasmid p12CC3$\_$GES-5 DNA (LC735983.1). Interestingly, this element was not selected as a mobile element by MobileElementFinder due to it being an uncharacterized plasmid within MobileElementFinder’s database of mobile elements. Additionally, this element was not selected as a mobile element by geNomad, possibly due to containing an uncharacterized nucleotide makeup pattern within the plasmid that was unspecific to any chromosome, plasmid, or viral marker. Thus, skandiver can efficiently identify novel mobile elements within whole genome assemblies without using gene annotation or markers. 

To demonstrate the speed and scalability of skandiver, we further attempted to verify whether the p12CC3$\_$GES-5 DNA plasmid discovered within a \textit{P. aeruginosa} whole genome assembly was present in any other assemblies or strains of \textit{P. aeruginosa}. We downloaded a set of 200 \textit{P. aeruginosa} complete genome assemblies from NCBI Assembly, and used skandiver to efficiently analyse all the assembly fragments. skandiver took about 20 minutes to process all 200 assemblies, with over 110,000 genome fragments. The assembly fragments were queried against the p12CC3$\_$GES-5 DNA plasmid using the baseline skani algorithm, which processed 138,800 \textit{P. aeruginosa} query sequences in approximately 0.19 seconds. From the results, it was determined that four unique \textit{P. aeruginosa} strains in the 200 assemblies queried (JAPEVK010000001.1, JAOVYS010000001.1, MPBS01000001.1, LOJK01000001.1) exhibited considerable matches of 95$\%$ ANI or higher against the reference plasmid. skandiver can not only identify novel mobile genetic elements but also rapidly search large datasets of MAGs for any strains that may share a mobile genetic element. 

\begin{table*}[htbp]
    \centering
    \caption{\textbf{skandiver, MobileElementFinder, and geNomad correctly do not identify known conserved elements as mobile.} Comparison of $>$10kbp conserved elements analyzed by different mobile element detection software. We curated a list of over two thousand conserved genes from NCBI was used. On the full list (including conserved genes $<$10kbp), skandiver had an overall false positive rate (FPR) of 2.2\%, geNomad 1.3\%, and MobileElementFinder 0\%. When we filtered down to only the five conserved genes above 10kbp in size, no false positives were found.
    }
    \label{tab:conservedpks}
    \resizebox{\textwidth}{!}{%
        \begin{tabular}{|c|c|c|c|c|c|c|c|}
            \hline
            \textbf{Protein ID} & \textbf{Species} & \textbf{Length (bp)} & \textbf{Genome Position} & \textbf{Function} & \textbf{skandiver} & \textbf{geNomad} & \textbf{MEFinder} \\
            \hline
            UFZ14059.1 & Streptomyces sp. & 13718 & 18513 & PKS I & N & N & N \\
            \hline
            UFZ14060.1 & Streptomyces sp. & 11240 & 32331 & PKS I & N & N & N \\
            \hline
            ABW96540.1 & Streptomyces spiroventicillatus & 32483 & 9369 & Polyketide synthase & N & N & N \\
            \hline
            ABW96541.1 & Streptomyces spiroventicillatus & 17132 & 26694 & Polyketide synthase & N & N & N \\
            \hline
            ABW96542.1 & Streptomyces spiroventicillatus & 17312 & 43846 & Polyketide synthase & N & N & N \\
            \hline
        \end{tabular}
        }
\end{table*}

\begin{figure}[b]
    \centering
    \includegraphics[width=0.9\linewidth]{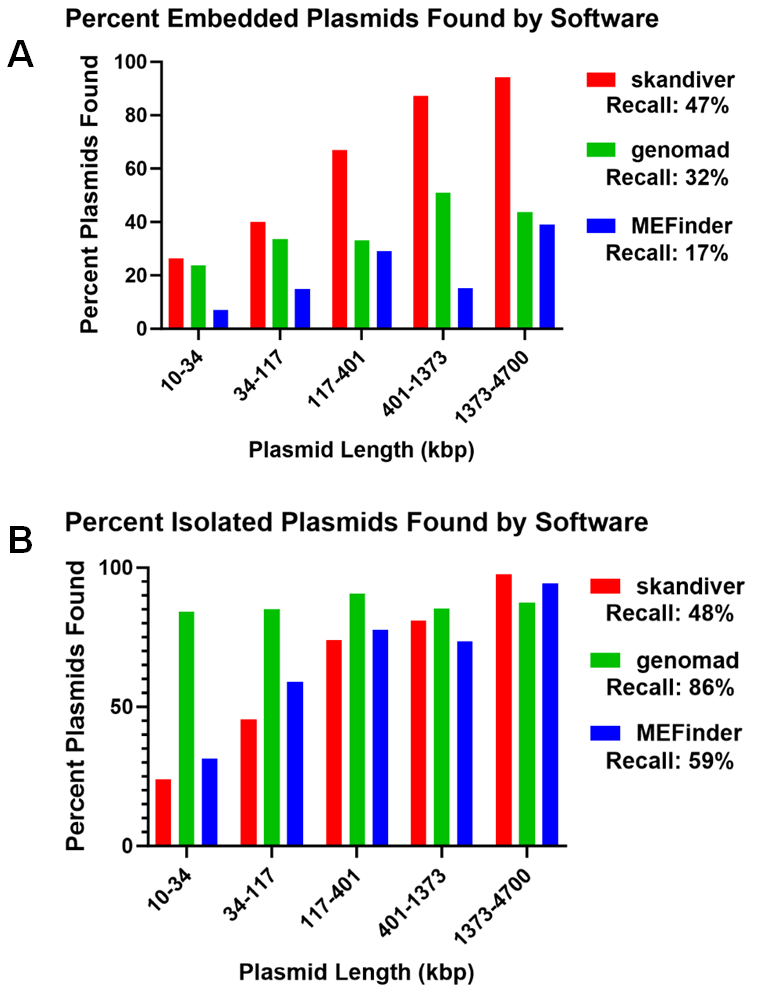}
    \caption{\textbf{Recall of mobile element finding software on known plasmid database.} Comparison of recall across skandiver, geNomad, and MobileElementFinder (MEFinder) on large plasmids ($>$10kbp) from NCBI. Recall was assessed for either \textbf{A.} plasmids embedded within a genome of \textit{E. coli} (CP008805.1) or \textbf{B.} isolated. For embedded plasmids, a database of large plasmids from NCBI was inserted into a genome (CP008805.1 Escherichia coli O157:H7 str. SS17, complete genome). Plasmids were placed 10kbp apart from each other, creating a large hybrid genome containing all plasmids embedded within a genome, thus representing a scenario where MGEs are integrated. This also ensured that the plasmid start location was not aligned with the 10kbp chunks used by skandiver. A plasmid was considered found by a method if that method labeled a region of the hybrid genome to be a MGE, and that region contained one of the inserted plasmids.}
    \label{fig:frankenplasmid}
\end{figure}

\begin{figure}[bt]
    \centering
    \includegraphics[width=1\linewidth]{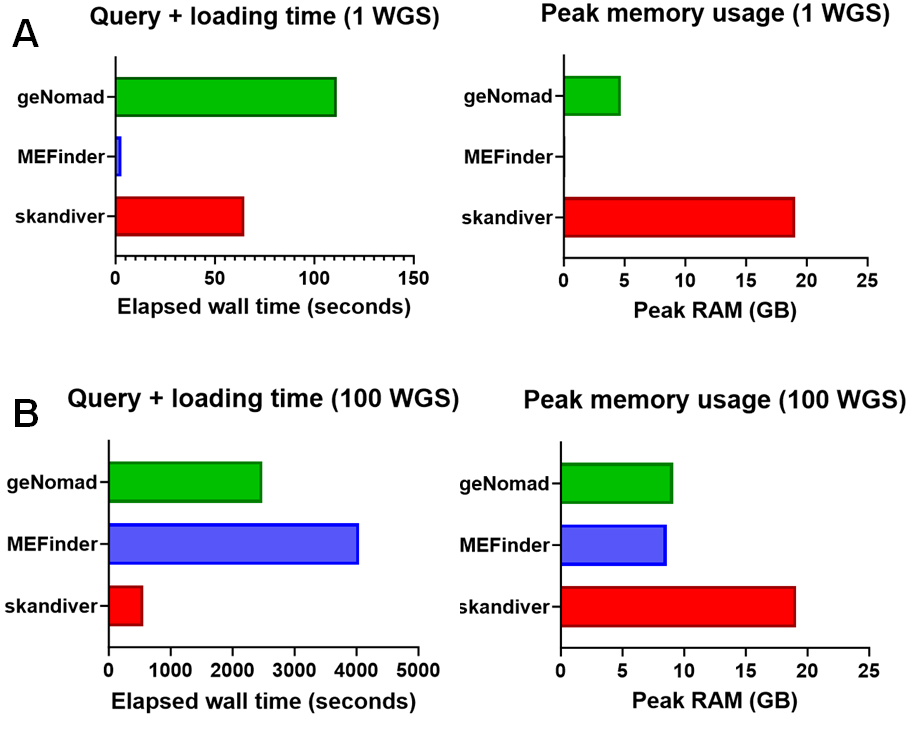}
    \caption{\textbf{skandiver scales well to larger datasets.} Comparison of skandiver against MobileElementFinder (MEFinder) and geNomad in terms of time (left column) and memory usage (right column) required for completing a query. skandiver, MEFinder, and geNomad were run on datasets consisting of \textbf{A.} one whole genome assembly and \textbf{B.} one hundred whole genome assemblies.}
    \label{fig:runtime}
\end{figure}

\subsection{Runtime, memory, and recall metrics}
skandiver can efficiently query 100 separate whole genome assemblies against the GTDB database of representative genomes ($>$65,000 bacteria species) in approximately 9 minutes (Figure \ref{fig:runtime}). skandiver is about 5 times faster when evaluating 100 whole genome assemblies simultaneously compared to MobileElementFinder and geNomad. While MobileElementFinder and geNomad use significantly less memory than skandiver for smaller sets of assemblies, this gap reduces for 100 whole genomes, as skandiver is better at scaling up, with a consistent memory footprint even as the scale of the analysis increases. This demonstrates skandiver’s scalability, with consistent speed and memory usage that allows researchers to analyze extensive datasets with speed and precision.

For completeness, we also extracted a list of large known plasmids ($>$10kbp) from NCBI (accession IDs available on Github repository), and then ran these sequences for detection of MGEs (Figure \ref{fig:frankenplasmid}). We considered skandiver to have correctly detected an MGE if one of its substring chunks was detected as mobile. Plasmids were either artificially inserted into a hypothetical genome of \textit{E. coli} (CP008805.1), or directly queried as independent isolates in the case of Figure \ref{fig:frankenplasmid}A and B respectively. Due to the extremely large size of the hypothetical genome ($>$1Gb), the genome was broken down into fragments of $~$5 million bp in size prior to analysis by geNomad. Notably, many of these plasmids are likely in the curated databases that geNomad and MobileElementFinder use for training/alignment, so it is not surprising that they have higher recalls in the case of isolated plasmids. However, skandiver still performs reasonably, especially for the largest plasmids, despite not using a curated database or a set of marker protein annotations.

In particular, skandiver exhibits significantly higher recalls when identifying plasmids embedded within a genome, which we believe is due to the other software using different detection methods for isolates versus embedded sequences (Figure \ref{fig:frankenplasmid}A). For metagenomic analysis, this may prove particularly useful when querying entire assembled genomes or contigs for potential mobile regions. In addition to this, skandiver's accuracy, while lower than MobileElementFinder and geNomad for isolated plasmids smaller than 1 million bp, demonstrates increasing accuracy as MGE size increases (Figure \ref{fig:frankenplasmid}B). skandiver plateaus around near-100\% accuracy for plasmids around 3 million bp in size, while the other tools plateau around 85\% accuracy. This suggests that skandiver may be particularly proficient in identifying large MGEs compared to existing software. 

To ensure skandiver reliably identifies mobile elements while minimizing false positives, we benchmarked skandiver against a list we curated of over two thousand conserved elements obtained from NCBI (accession IDs available on Github repository). Our results showed that skandiver had an overall FPR of 2.2\%, while geNomad had a lower FPR of 1.3\%, and MobileElementFinder achieved an FPR of 0\%. When filtering our conserved element dataset to only include elements above 10kbp in size, none of the methods produced false positives (Table \ref{tab:conservedpks}). As such, skandiver's ability to avoid false positives is comparable to existing mobile element detection software, especially in the $>$10kbp regime this manuscript works in.

\section{Conclusion}

In this manuscript, we argued in favor of an evolutionary-divergence based characterization of mobile-genetic elements, which bypasses some of the limitations of existing bioinformatics tools like geNomad and MobileElementFinder, which rely on reference databases (either through training or alignment respectively). Our new tool, skandiver, is a scalable and versatile approach for the identification of potential mobile genetic elements within whole genome assemblies that can detect novel mobile genetic elements.
We have shown that skandiver is efficient enough to be run on hundreds of genomes and provides an orthogonal way of finding large putative MGEs that does not require annotation or training data. skandiver excels in scenarios involving high ANI values ($>$90$\%$) and comparisons against diverse genomic datasets, making it especially suited for finding the distinct evolutionary patterns reflective of mobile elements. 

However, we also showed that there is very little concordance on real data between methods for finding mobile genetic elements. All three methods are indeed finding mobile genetic elements, but none of the methods have particularly high sensitivity. Given that all of these methods have reasonable runtimes, it would not be an unexpected workflow for a practitioner to use geNomad or MobileElementFinder to find known MGEs, and then use skandiver to try to find some of the remaining uncharacterized MGEs. Ultimately, despite all the progress made in recent years, it seems that our ability as a bioinformatics community to comprehensively characterize mobile genetic elements at scale is still incomplete, and there exists substantial scope for further scientific exploration in this domain.

Future directions for skandiver in particular could include refining parameters and exploring methodologies to allow skandiver to distinguish different classes of 
 (especially smaller) mobile genetic elements (insertion sequences, transposases, prophages, integrative conjugative elements, etc.). 
Furthermore, skandiver's ability to efficiently requery large datasets for previously discovered mobile elements may allow it to construct a profile of the preferred mobile target genomes of mobile elements.
Another way to leverage efficiently querying large datasets is to use a pangenome of strain-specific genomes, rather than just the GTDB. This may allow for increased sensitivity. One of the difficulties is that the TimeTree of Life may not include evolutionary divergence times between those genomes, but it may be possible to either compute those directly or use ANI as a rough proxy
More generally though, we hope that future mobile genetic element detectors will incorporate our divergence-based metric with alignment and machine learning models to build a more comprehensive understanding of MGEs.

\section{Acknowledgments}
We acknowledge funding from the Natural Sciences and Engineering Research Council of Canada (NSERC) grant RGPIN-2022-03074, as well as startup funding from Carnegie Mellon University. J.S was supported by an NSERC CGS-D scholarship.

\bibliographystyle{plainnat}
\bibliography{reference}

\begin{thebibliography}{20}
\providecommand{\natexlab}[1]{#1}
\providecommand{\url}[1]{\texttt{#1}}
\expandafter\ifx\csname urlstyle\endcsname\relax
  \providecommand{\doi}[1]{doi: #1}\else
  \providecommand{\doi}{doi: \begingroup \urlstyle{rm}\Url}\fi

\bibitem[Altschul et~al.(1990)Altschul, Gish, Miller, Myers, and Lipman]{altschul1990basic}
Stephen~F Altschul, Warren Gish, Webb Miller, Eugene~W Myers, and David~J Lipman.
\newblock Basic local alignment search tool.
\newblock \emph{Journal of molecular biology}, 215\penalty0 (3):\penalty0 403--410, 1990.

\bibitem[Berger and Yu(2023)]{berger2023navigating}
Bonnie Berger and Yun~William Yu.
\newblock Navigating bottlenecks and trade-offs in genomic data analysis.
\newblock \emph{Nature Reviews Genetics}, 24\penalty0 (4):\penalty0 235--250, 2023.

\bibitem[Camargo et~al.(2023)Camargo, Roux, Schulz, Babinski, Xu, Hu, Chain, Nayfach, and Kyrpides]{camargo2023}
Antonio~Pedro Camargo, Simon Roux, Frederik Schulz, Michal Babinski, Yan Xu, Bin Hu, Patrick~SG Chain, Stephen Nayfach, and Nikos~C Kyrpides.
\newblock Identification of mobile genetic elements with genomad.
\newblock \emph{Nature Biotechnology}, pages 1--10, 2023.

\bibitem[Camargo et~al.(2024)Camargo, Call, Roux, Nayfach, Huntemann, Palaniappan, Ratner, Chu, Mukherjeep, Reddy, et~al.]{camargo2024mode}
Antonio~Pedro Camargo, Lee Call, Simon Roux, Stephen Nayfach, Marcel Huntemann, Krishnaveni Palaniappan, Anna Ratner, Ken Chu, Supratim Mukherjeep, TBK Reddy, et~al.
\newblock Img/pr: a database of plasmids from genomes and metagenomes with rich annotations and metadata.
\newblock \emph{Nucleic acids research}, 52\penalty0 (D1):\penalty0 D164--D173, 2024.

\bibitem[Cock et~al.(2009)Cock, Antao, Chang, Chapman, Cox, Dalke, Friedberg, Hamelryck, Kauff, Wilczynski, et~al.]{cock2009biopython}
Peter~JA Cock, Tiago Antao, Jeffrey~T Chang, Brad~A Chapman, Cymon~J Cox, Andrew Dalke, Iddo Friedberg, Thomas Hamelryck, Frank Kauff, Bartek Wilczynski, et~al.
\newblock Biopython: freely available python tools for computational molecular biology and bioinformatics.
\newblock \emph{Bioinformatics}, 25\penalty0 (11):\penalty0 1422, 2009.

\bibitem[Davies and Davies(2010)]{davies2010}
Julian Davies and Dorothy Davies.
\newblock Origins and evolution of antibiotic resistance.
\newblock \emph{Microbiology and molecular biology reviews}, 74\penalty0 (3):\penalty0 417--433, 2010.

\bibitem[Doolittle et~al.(1996)Doolittle, Feng, Tsang, Cho, and Little]{doolittle1996determining}
Russell~F Doolittle, Da-Fei Feng, Simon Tsang, Glen Cho, and Elizabeth Little.
\newblock Determining divergence times of the major kingdoms of living organisms with a protein clock.
\newblock \emph{Science}, 271\penalty0 (5248):\penalty0 470--477, 1996.

\bibitem[Durrant et~al.(2020)Durrant, Li, Siranosian, Montgomery, and Bhatt]{durrant2020}
Matthew~G Durrant, Michelle~M Li, Benjamin~A Siranosian, Stephen~B Montgomery, and Ami~S Bhatt.
\newblock A bioinformatic analysis of integrative mobile genetic elements highlights their role in bacterial adaptation.
\newblock \emph{Cell host \& microbe}, 27\penalty0 (1):\penalty0 140--153, 2020.

\bibitem[Forster et~al.(2022)Forster, Liu, Kumar, Gulliver, Gould, Escobar-Zepeda, Mkandawire, Pike, Shao, Stares, et~al.]{forster2022}
Samuel~C Forster, Junyan Liu, Nitin Kumar, Emily~L Gulliver, Jodee~A Gould, Alejandra Escobar-Zepeda, Tapoka Mkandawire, Lindsay~J Pike, Yan Shao, Mark~D Stares, et~al.
\newblock Strain-level characterization of broad host range mobile genetic elements transferring antibiotic resistance from the human microbiome.
\newblock \emph{Nature Communications}, 13\penalty0 (1):\penalty0 1445, 2022.

\bibitem[Frost et~al.(2005)Frost, Leplae, Summers, and Toussaint]{frost2005mobile}
Laura~S Frost, Raphael Leplae, Anne~O Summers, and Ariane Toussaint.
\newblock Mobile genetic elements: the agents of open source evolution.
\newblock \emph{Nature Reviews Microbiology}, 3\penalty0 (9):\penalty0 722--732, 2005.

\bibitem[Johansson et~al.(2021)Johansson, Bortolaia, Tansirichaiya, Aarestrup, Roberts, and Petersen]{johansson2021}
Markus~HK Johansson, Valeria Bortolaia, Supathep Tansirichaiya, Frank~M Aarestrup, Adam~P Roberts, and Thomas~N Petersen.
\newblock Detection of mobile genetic elements associated with antibiotic resistance in salmonella enterica using a newly developed web tool: Mobileelementfinder.
\newblock \emph{Journal of Antimicrobial Chemotherapy}, 76\penalty0 (1):\penalty0 101--109, 2021.

\bibitem[Khedkar et~al.(2022)Khedkar, Smyshlyaev, Letunic, Maistrenko, Coelho, Orakov, Forslund, Hildebrand, Luetge, Schmidt, et~al.]{khedkar2022}
Supriya Khedkar, Georgy Smyshlyaev, Ivica Letunic, Oleksandr~M Maistrenko, Luis~Pedro Coelho, Askarbek Orakov, Sofia~K Forslund, Falk Hildebrand, Mechthild Luetge, Thomas~SB Schmidt, et~al.
\newblock Landscape of mobile genetic elements and their antibiotic resistance cargo in prokaryotic genomes.
\newblock \emph{Nucleic acids research}, 50\penalty0 (6):\penalty0 3155--3168, 2022.

\bibitem[Kumar et~al.(2022)Kumar, Suleski, Craig, Kasprowicz, Sanderford, Li, Stecher, and Hedges]{kumar2022}
Sudhir Kumar, Michael Suleski, Jack~M Craig, Adrienne~E Kasprowicz, Maxwell Sanderford, Michael Li, Glen Stecher, and S~Blair Hedges.
\newblock Timetree 5: an expanded resource for species divergence times.
\newblock \emph{Molecular Biology and Evolution}, 39\penalty0 (8):\penalty0 msac174, 2022.

\bibitem[Paez-Espino et~al.(2016)Paez-Espino, Chen, Palaniappan, Ratner, Chu, Szeto, Pillay, Huang, Markowitz, Nielsen, et~al.]{paez2016img}
David Paez-Espino, I-Min~A Chen, Krishna Palaniappan, Anna Ratner, Ken Chu, Ernest Szeto, Manoj Pillay, Jinghua Huang, Victor~M Markowitz, Torben Nielsen, et~al.
\newblock Img/vr: a database of cultured and uncultured dna viruses and retroviruses.
\newblock \emph{Nucleic acids research}, 45\penalty0 (D1):\penalty0 gkw1030, 2016.

\bibitem[Parks et~al.(2022)Parks, Chuvochina, Rinke, Mussig, Chaumeil, and Hugenholtz]{parks2022gtdb}
Donovan~H Parks, Maria Chuvochina, Christian Rinke, Aaron~J Mussig, Pierre-Alain Chaumeil, and Philip Hugenholtz.
\newblock Gtdb: an ongoing census of bacterial and archaeal diversity through a phylogenetically consistent, rank normalized and complete genome-based taxonomy.
\newblock \emph{Nucleic acids research}, 50\penalty0 (D1):\penalty0 D785--D794, 2022.

\bibitem[Partridge et~al.(2018)Partridge, Kwong, Firth, and Jensen]{partridge2018mobile}
Sally~R Partridge, Stephen~M Kwong, Neville Firth, and Slade~O Jensen.
\newblock Mobile genetic elements associated with antimicrobial resistance.
\newblock \emph{Clinical microbiology reviews}, 31\penalty0 (4):\penalty0 10--1128, 2018.

\bibitem[Roux et~al.(2019)Roux, Adriaenssens, Dutilh, Koonin, Kropinski, Krupovic, Kuhn, Lavigne, Brister, Varsani, et~al.]{roux2019}
Simon Roux, Evelien~M Adriaenssens, Bas~E Dutilh, Eugene~V Koonin, Andrew~M Kropinski, Mart Krupovic, Jens~H Kuhn, Rob Lavigne, J~Rodney Brister, Arvind Varsani, et~al.
\newblock Minimum information about an uncultivated virus genome (miuvig).
\newblock \emph{Nature biotechnology}, 37\penalty0 (1):\penalty0 29--37, 2019.

\bibitem[Shapiro(2012)]{shapiro2012mobile}
James Shapiro.
\newblock \emph{Mobile genetic elements}.
\newblock Elsevier, 2012.

\bibitem[Shaw and Yu(2023)]{shaw2023}
Jim Shaw and Yun~William Yu.
\newblock Fast and robust metagenomic sequence comparison through sparse chaining with skani.
\newblock \emph{Nat Methods}, 22:\penalty0 1661–1665, 2023.

\bibitem[Wheeler et~al.(2007)Wheeler, Barrett, Benson, Bryant, Canese, Chetvernin, Church, DiCuccio, Edgar, Federhen, et~al.]{wheeler2007database}
David~L Wheeler, Tanya Barrett, Dennis~A Benson, Stephen~H Bryant, Kathi Canese, Vyacheslav Chetvernin, Deanna~M Church, Michael DiCuccio, Ron Edgar, Scott Federhen, et~al.
\newblock Database resources of the national center for biotechnology information.
\newblock \emph{Nucleic acids research}, 36\penalty0 (suppl\_1):\penalty0 D13--D21, 2007.

\end{thebibliography}

\clearpage
\newpage
\onecolumn
\section{Supplemental Figures}
\renewcommand{\figurename}{Supp Fig.}
\setcounter{figure}{0} 
\begin{figure}[h!]
    \centering
    \includegraphics[width=1\textwidth,trim={0 100px 0 0},clip]{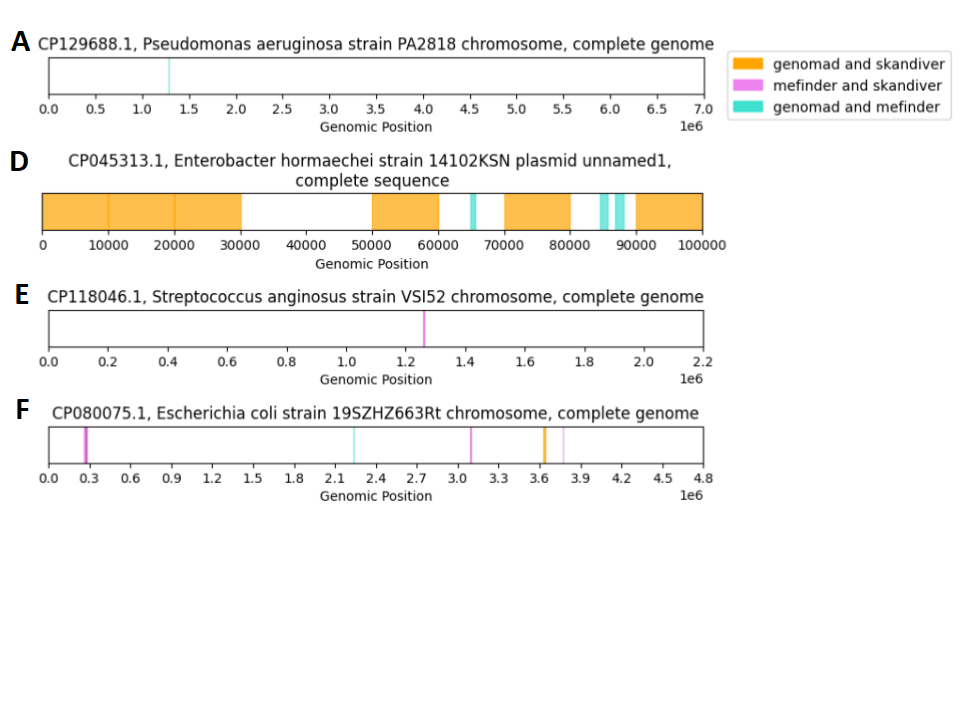}
    \caption{\textbf{Overlap of mobile element finding profiles of skandiver, MobileElementFinder (MEFinder), and geNomad: the three tools find different sets of putative genetic mobile elements.} Building off of Figure 3, we labeled areas where skandiver, MobileElementFinder, and geNomad found overlapping regions corresponding to potential mobile elements. Three of the seven genome assemblies shown in Figure 3 did not contain any regions of overlap between the three methods, and are not shown here;
    \textbf{A.} \textit{Pseudomonas aeruginosa} strain PA2818 chromosome,  \textbf{D. }\textit{Enterobacter hormaechei} strain 14102KSN plasmid, \textbf{E. }\textit{Streptococcus anginosus} strain VS152 chromosome, \textbf{F. }\textit{Escherichia coli} strain 19SZHZ663Rt chromosome. }
    \label{fig:comparison-overlap}
\end{figure}

\end{document}